\documentclass[epj,final]{svjour}

\usepackage{array}
\usepackage{rotating}
\usepackage{graphicx}
\usepackage{enumerate}
\usepackage{psfrag}
\usepackage{amssymb}
\usepackage{cite}

\def\abs#1{\left \vert\, #1 \right \vert}

\def\sech{\mathrm{sech}}

\newcounter{eq_abc}
\newcommand{\eqin}{\setcounter{eq_abc}{\value{equation}}%
  \stepcounter{eq_abc}\setcounter{equation}{0}%
  \renewcommand{\theequation}{\mbox{\arabic{eq_abc}\alph{equation}}}}%
\newcommand{\eqout}{\setcounter{equation}{\value{eq_abc}}%
  \renewcommand{\theequation}{\arabic{equation}}}%

\begin{document}

\title{Assessment of interspecies scattering lengths $a_{12}$ from stability
       of  two-component Bose-Einstein condensates}
\titlerunning{Assessment of interspecies scattering lengths $a_{12}$ \dots}
\authorrunning{Barnab{\'{a}}s Apagyi et al.}

\author{Barnab{\'{a}}s Apagyi \and D{\'{a}}niel Schumayer}
\institute{Institute of Physics,
         Budapest University of Technology and Economics,
         H-1111, Budafoki {\'{u}}t 8, Hungary \\
         \email{apagyi@phy.bme.hu, schumayer@phy.bme.hu}}

\abstract{A stability method is used  to assess possible values of
interspecies scattering lengths $a_{12}$ in two-component
Bose-Einstein condensates described within the Gross-Pitaevskii
approximation. The technique, based on a recent stability analysis
of solitonic excitations in two-component Bose-Einstein
condensates, is applied to ninety combinations of atomic alkali
pairs with given singlet and triplet intraspecies scattering
lengths as input parameters. Results obtained for values of
$a_{12}$ are in a reasonable agreement with the few ones available
in the literature and with those obtained from a Painlev{\'e}
analysis of the coupled Gross-Pitaevskii equations.}

\maketitle

\section{Introduction}
\label{sec:introduction}

Over the past decade an intensive interest is devoted to atomic and
molecule Bose-Einstein condensation. Recent progress in experimental
techniques makes it possible in the nearest future to engineer
two-component Bose-Einstein condensates (BECs) that are composed of
two different alkali-metal atoms. Until now, merely theoretical
studies have been carried out for two-component alkali-metal atom BEC
systems like Li-Rb~\cite{Apagyi2004}, Na-Rb \cite{Apagyi2004, Ho1996,
Pu1998_1, Pu1998_2}, and K-Rb \cite{Ferrari2002, Simoni2003}.
However, the mixture Cs-Li \cite{Mosk2001, Mudrich2002} has been
investigated experimentally, without reaching the BEC phase.

One of the most important quantities in engineering BECs is the
scattering length $a_{ij}$ characterizing the strength of the
interaction between the atoms of types $i$ and $j$. A realization of
two-component BECs depends on three scattering lengths, namely the
intraspecies scattering lengths $a_{11}$, $a_{22}$ and the
interspecies ones $a_{12}=a_{21}$. Accurate knowledge of these three
quantities is of great importance in design and implementation of
two-component BECs. However, the problem is that the scattering
lengths between many like alkali-metal atoms are known, whereas those
between unlike alkalis have not yet been measured or calculated,
except for very few cases (e.g. \cite{SL39_41K, Burke1999, Mosk2001,
Ferrari2002, Simoni2003, Jamieson2003}).

The aim of this note is to apply a simple stability method to assess
possible values of interspecies scattering length by using a recent
analysis \cite{Apagyi2004} for existence of solitons within the
two-component BEC. The basic idea of the formulation is that the
solitons are stable objects and if they can be created in the
background BEC matter then the latter should show up also stability
property. Various types of solitonic excitations can be created in
two-component BEC matter, such as those of bright (B) and dark (D)
type or of BB and DD types. Solitonic excitations in BEC matter is
dealt with by many authors including \cite{Dum1998a, Denschlag2000,
Busch2001a, Ohberg2001a, Khaykovich2002a, Kevrekidis2003}. In this
work we shall treat only BD type excitations which provide finite
interval for possible values of interspecies scattering length
$a_{12}$.

We shall determine, in tabulated form, those ranges of $a_{12}$ for
which various two-component BECs maintain stability with respect to
the solitonic excitations. The background is treated in the
Thomas-Fermi (TF) approximation while the excitations are considered
in the form of static BD solitons. Furthermore, we restrict the
treatment of the dynamics to one dimension, but employ
three-dimensional scattering lengths using an appropriate crossing
area $A$. Note that cigar-like quasi one-dimensional BECs have
already been produced by micro-trap devices developed quite recently
\cite{Fortagh1998, Ott2001, Hansel2001a, Leanhardt2002}. As a case
study we shall treat in detail two specific BEC systems, namely those
composed of the alkali-metal atom pairs K-Rb and Li-Na. For these
systems we exhibit the BD solitonic density profiles at the most
probable values of $a_{12}$ which ensure integrability (i.e. inverse
scattering solution of the Gross-Pitaevksii equations), as well as
the allowable range of $a_{12}$ which takes care of particle number
conservation.

The results of the present investigation indicate that given the
intraspecies scattering length $a_{ii}$, $i=1,2$ the method is
capable to provide possible values for the interspecies scattering
length $a_{12}$, at which the two-component BEC may prove stable.
Whether or not the true (physical) values of $a_{12}$ fall into these
stability ranges is a question. If yes, then one can further
calculate for finer details of the two-component BEC, as for the
particle numbers $N_{i}$ ($i=1,2$), the extension of components and
so on. If not, then it would mean that other types (e.g. BB or DD) of
solitonic excitations compatible with the physical value of $a_{12}$
could be realized. Moreover, one may try to adjust $a_{ij}$'s
slightly by changing the strength of the magnetic field
\cite{Cubizolles2003}, to meet the stability requirement formulated
in the present work. (Note, however, that the Feshbach resonance
method is not fully developed for interatomic collisions.) A third
possibility is to choose other alkali pair being more favorable from
point of view of values of $a_{ij}$'s for assembling the
two-component BEC satisfying stability condition formulas given in
the present method.

In section \ref{sec:TheoreticalBackground} the theoretical background
will be outlined. Section \ref{sec:results} contains the estimated
ranges of interspecies scattering length $a_{12}$, in tabulated form,
for which the solitonic stability requirement holds and a comparison
with earlier findings and those resulted from Painlev{\'e} analysis
will be made. Section \ref{sec:summary_of_the_method} is devoted to a
short summary of the method.

\section{Theoretical background}
  \label{sec:TheoreticalBackground}

Following \cite{Apagyi2004} we write the coupled Gross-Pitaevskii
(GP) equations \cite{Gross1961, Pitaevskii1961} describing a quasi
one-dimensional two-component BEC in the form ($i=1,2$):
\begin{equation} \label{eq:coupled_GP_math_notation}
   i \hbar \psi_{i,t} = \left \lbrack - {\hbar^2\over 2m_{i}} \, \partial_{xx} +
                           \sum_{j=1}^{2}{\Omega_{ij} \abs{\psi_{j}}^{2}} + V_{i}
                           \right \rbrack \! \psi_{i}
\end{equation}
where $m_{i}$ denotes the individual mass of the $i$th atomic
species, $\Omega_{ij} = 2\pi\hbar^2 a_{ij} /A\mu_{ij}$ with $a_{ij}$
being the 3D scattering length, $A$ represents a general transverse
crossing area of the cigar-shape BEC, $\mu_{ij}=m_{i} m_{j}/(m_{i} +
m_{j})$ is the reduced mass, and $V_{i}$, ($i=1,2$) denotes the
external trapping potentials. In the case of real trap potentials the
normalization of the wave functions reads as $N_{i} =
\int_{-\infty}^{\infty}{\!\! \abs{\psi_{i}}^{2}\!dx}$, $(i=1,2)$ with
$N_{i}$ denoting the number of atoms in the $i$th component. We do
not allow for the species to transform into each other, so that the
numbers of particles are conserved quantities for both alkalies.

The stationary solutions of the coupled GP equation
(\ref{eq:coupled_GP_math_notation}) can be written in the form
$\psi_{i}(x,t) = \Phi_{i}(x) \exp(-iE_{i}t/ \hbar)$ where $E_{i}$
represents the one-particle energy of the $i-$th component particle
($i=1,2$). Using this formula and neglecting the kinetic terms one is
able to derive the approximate density profiles and a semi-infinite
range for $a_{12}$ in which the Thomas-Fermi ground state exists. In
the case of a harmonic trap potential the calculation results in the
usual inverse-parabolic distribution.

We then look for the excited static solutions in the form
$\tilde{\psi_{i}}(x,t) = \Phi_{i}(x) \phi_{i}(x)
\exp(-i \tilde{E}_it /\hbar)$ with $\phi_{i}(x)$ being an excess or
defect of the $i-$th component of the background density, triggered
by an excitation mechanism. (Note that such trial function for a
solitonic excitation has been applied earlier by Dum et
al~\cite{Dum1998a}.) Inserting the above ansatz into the GP equations
(\ref{eq:coupled_GP_math_notation}) and assuming that the excitation
mechanism restricts the change of density only to a small interval
around $x=0$ where the TF solutions can be approximated by
$\Phi_{i}(0)$, the following coupled equations can be obtained for
the perturbing functions ($i=1,2$)
\begin{equation} \label{eq:coupled_GP_soliton}
   \tilde{E_{i}} \phi_{i} = \left \lbrack- \frac{\hbar^{2}}{2m_{i}} \, \partial_{xx} +
                            \sum_{j=1}^{2}{\tilde{\Omega}_{ij} \abs{\phi_{j}}^{2}}
                            \right \rbrack \phi_{i}
\end{equation}
where $\tilde{\Omega}_{ij} \propto \Omega_{ij}$ ($i,j=1,2$) and the
very small potential terms have been neglected.

The coupled equations above can be solved for the perturbing
functions $\phi_{i}(x)$ by using the static bright-dark (BD) coupled
solitonic ansatz
 \eqin
 \begin{eqnarray} \label{eq:ansatz}
    \phi_{B1} (x) &=& q_{1} \sech{(k x)}, \hspace{6mm}
    \phi_{B1}(x \to \pm \infty)=0,
    \\
    \phi_{D2} (x) &=& q_{2} \tanh{(k x)},  \hspace{4mm}
    \phi_{D2}(x \to \pm \infty) = \pm q_{2},
 \end{eqnarray}
\eqout with the specified boundary conditions, generic complex
amplitudes $q_{i}$, and common real size parameter $k$.

The requirement that the moduli of the two amplitudes $q_{1}$ and
$q_{2}$ be positive real numbers gives the stability conditions
$(A_{ij}=a_{ij}(1+m_{i}/m_{j}))$
 \eqin
 \begin{eqnarray} \label{eq:ranges}
    f_{B1}(a_{12}) &=& \frac{A_{12}-A_{22}}{\det(A)} \ge 0,
    \\
    f_{D2}(a_{12}) &=& \frac{A_{11}-A_{21}}{\det(A)} \ge 0
 \end{eqnarray}
 \eqout
for the existence of BD solitonic excitation within the two-component
BEC. The finite ranges of $a_{12}$ provided by the solitonic
stability requirement (\ref{eq:ranges}-b) are naturally narrower than
the TF domain of $a_{12}$ prescribed by the positivity of the density
$\abs{\Phi_{i}(x)}^{2}$.

Let us note here that instead of the BD solitonic ansatz
(\ref{eq:ansatz}-b) one might as well apply BB or DD prescriptions
which would still satisfy equations (\ref{eq:coupled_GP_soliton}).
These ansatzes result in similar stability conditions like the forms
(\ref{eq:ranges}-b) but yield different stability domain of $a_{12}$.
This is one reason for the usage of the terminology
{\emph{assessment}} by which we mean that the physical value of
$a_{12}$ may be well outside the ranges given later in this paper.
Nevertheless, for the specific conditions corresponding to the BD
stability we shall find compatibility between our findings and
others, including the P-test $a_{12}$ values which are obtained from
the singularity structure analysis of the coupled GP equations.

From the many possibilities we shall select two examples, with alkali
pairs of Li-Na and K-Rb, representing realistic possibilities for
future experimental study.

\subsection[Static BD solitonic excitations \dots]%
           {Static BD solitonic excitations of Li-Na and K-Rb condensates%
            \label{subsec:static_solitonic_excitations}}

\begin{figure}[htb!]
    \begin{center}
    \psfrag{[wm]}{\footnotesize{$[\mu \mathrm{m}]$}}
    \psfrag{[mum1]}{$[\mu \mathrm{m}^{-1}]$}
    \psfrag{phi}{\footnotesize{$\vert\tilde{\psi}_{i} (x)\vert^{2}$}}
    \includegraphics[width=40mm]{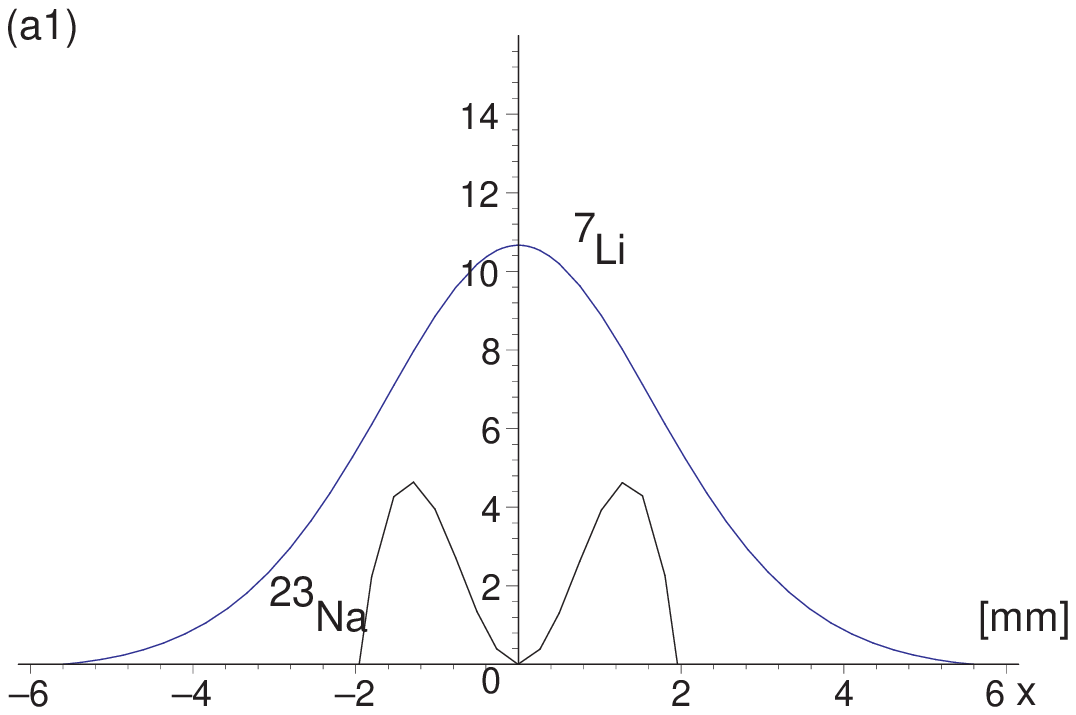}
    \hspace*{2mm}
    \includegraphics[width=40mm]{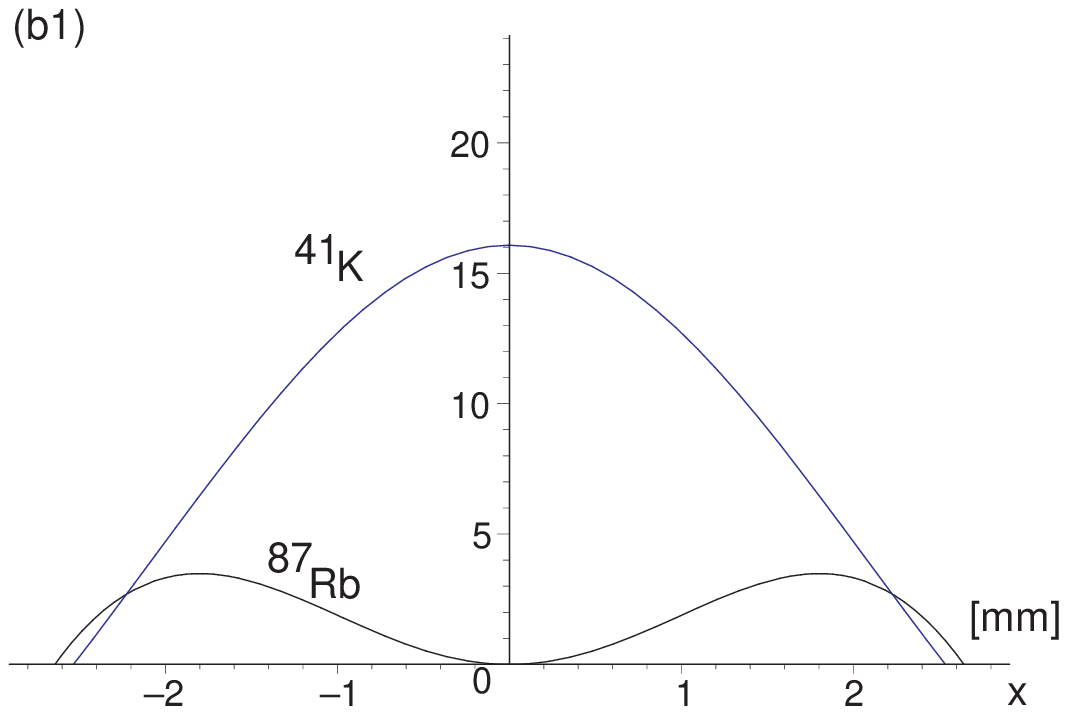}\newline
    \includegraphics[width=40mm]{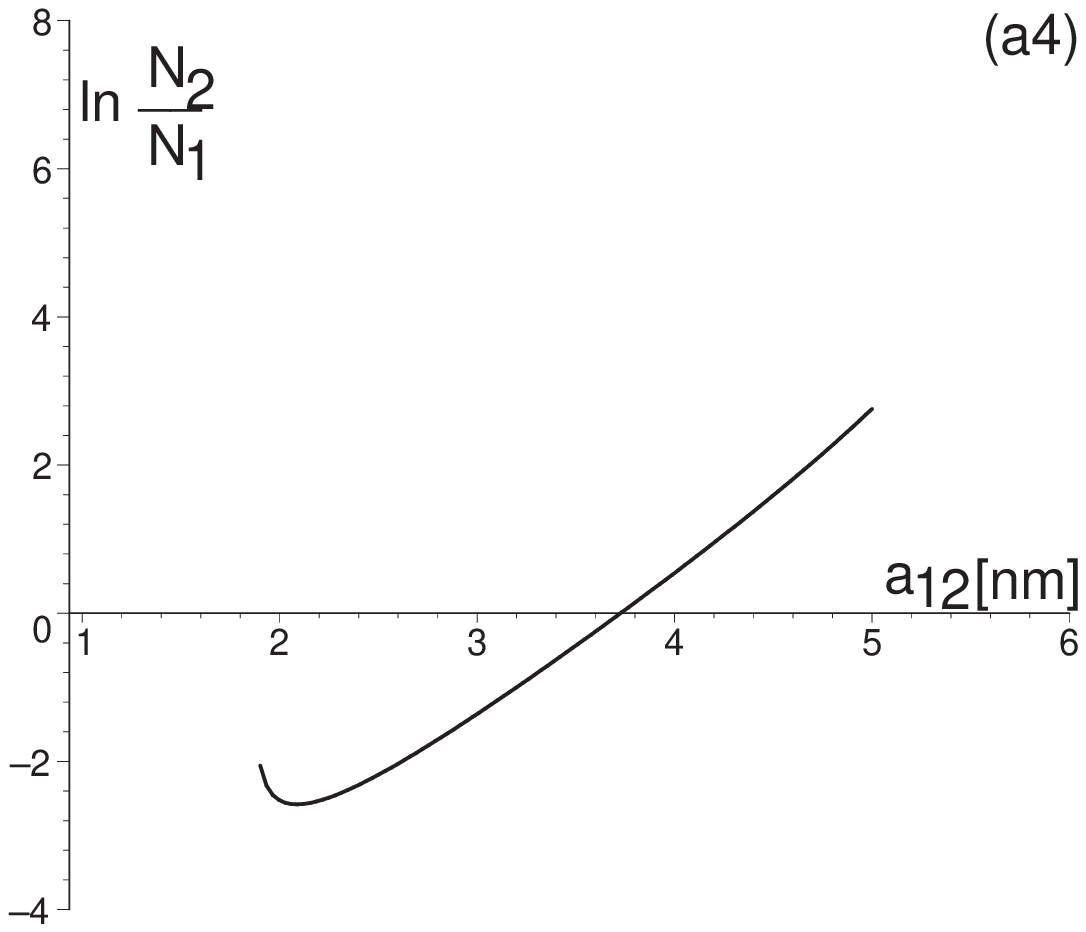}
    \hspace*{3mm}
    \includegraphics[width=40mm]{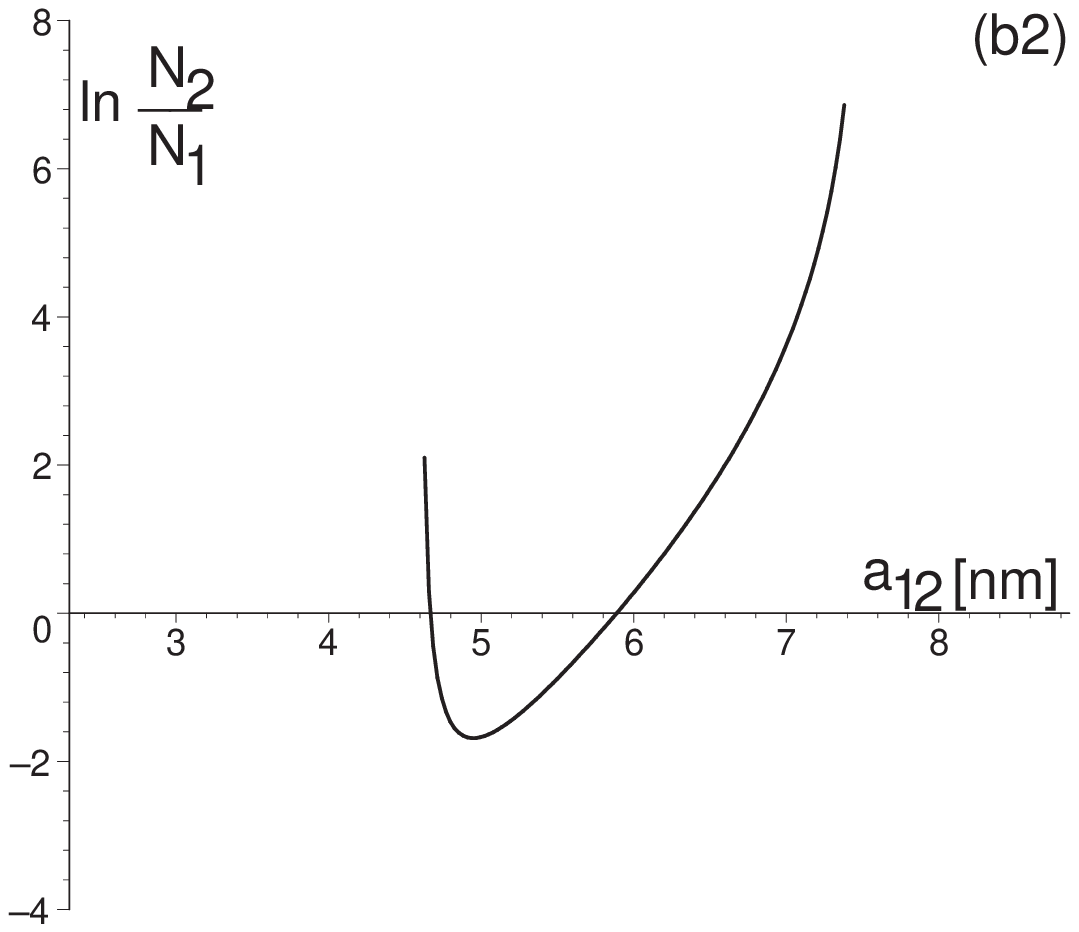}\newline
    \caption{Absolute squares of wave functions,
             $\vert \tilde{\psi_{1}} \vert^{2}$ and $\vert \tilde{\psi_{2}}
             \vert^{2}$ in unit [$\mu m^{-1}$], of the BD solution of the
             coupled Gross-Pitaevskii equation~(\ref{eq:coupled_GP_math_notation})
             as a function of the position $x$ for different BEC systems.
             (a) triplet $^{7}$Li-$^{23}$Na system: $N_{2}=10^{4}$,
                 $N_{1}\approx 4.7 \times 10^{4}$,
                 $a_{11}=-1.4$\, nm ($^7$Li), $a_{22}=4.0$\, nm ($^{23}$Na)
                 $a_{12}= 2.90$ nm, trap frequencies are:
                 $\omega_{1} = 2\pi\times 530$\,Hz,
                 $\omega_{2} = 2\pi \times 292$\,Hz.
             (b) triplet $^{41}$K-$^{87}$Rb system: $N_{2}=10^{4}$,
                 $N_{1}\approx 4.9 \times 10^{4}$,
                 $a_{11}=3.4$\, nm ($^{41}$K), $a_{22}=5.5$\, nm ($^{87}$Rb),
                 $a_{12}= 4.85$ nm, trap frequencies are:
                 $\omega_{1} = 2\pi \times 220$\,Hz,
                 $\omega_{2} = 2\pi\times 150$\,Hz.
             Figures (a2) and (b2) show the dependence of the logarithmic ratio
             $\ln{\!\left (N_{2}/N_{1} \right )}$ on the interspecies scattering
             length $a_{12}$ for these two-component BECs.
             The transverse crossing area $A=r^{2} \pi$ with radius $r = 0.3$ nm has
             been employed throughout.
             \label{fig:bd_amplitudes}}
    \end{center}
\end{figure}

In order to illustrate the solitonic excitations we show in figure
\ref{fig:bd_amplitudes} the density profiles $\vert
\tilde{\psi}_{i}(x) \vert^{2}$ of the two BEC systems composed of
lithium-sodium and potassium-rubidium atoms. All of the components
are chosen to be in the triplet scattering state. We have kept
$N_{2}$ fixed by choosing the dark soliton particle number to be
$N_{2}=10^{4}$ for both the $^{7}$Li-$^{23}$Na triplet system and
the $^{41}$K-$^{87}$Rb two-component condensate. The bright
component has particle number $N_{1} \approx 5 \times N_{2}$ in
both cases, corresponding to the chosen values of $a_{12}=2.9$ nm
(Li-Na) and $a_{12}=4.85$ nm (K-Rb). The excitation of the BEC is
manifested by the appearance of the B and D solitons on the TF
background density.

If the system parameters $m_{1}$, $m_{2}$, $a_{11}$ and $a_{22}$ are
given, then one can determine the broadest range of the interspecies
scattering length by using the inequalities (\ref{eq:ranges}-b). This
procedure does not involve the particle numbers $N_{1}$ and $N_{2}$
and gives an assessment of $a_{12}$ range for which the existence of
BD or DB solitons in the two-component BEC might be expected.
Particle number conservation further restricts the interval of
$a_{12}$ and leads to a system coupled equations for $N_{2}/N_{1}$
and the size parameter $k$ (see eqs. (11) of Ref. \cite{Apagyi2004}).
Solving these equations, the final results are shown in figures
\ref{fig:bd_amplitudes}(a2) and \ref{fig:bd_amplitudes}(b2). We
observe here a rather smooth behavior of $N_{2}/N_{1}$, as well as a
shrinking of the allowable intervals from $a_{12}=-0.6$; $6.1$ nm (as
determined by inequalities~(\ref{eq:ranges}-b)) to $a_{12}=1.9$;
$4.9$ nm (obtained by use of particle number conservation) for the
$^{7}$Li-$^{23}$Na system, and from $a_{12}=4.1$; $7.5$ nm to
$a_{12}=4.5$; $7.4$ nm for the $^{41}$K-$^{87}$Rb configuration.

In conclusion we can say that if the actual value of $a_{12}$
determined by some experimental or theoretical method turns out to
fall within the range determined by inequalities (\ref{eq:ranges}-b)
then one may fine tune the particle number ratio $N_{2}/N_{1}$ and
the size parameter $k$. In this way we get complete information about
both the background and solitonic stability.

\subsection{$a_{12}$ values resulting from Painlev{\'e} analysis%
            \label{subsec:a12_from_Painleve}}

In Ref.~\cite{Apagyi2001} it has been shown that the coupled GP
equations pass the Painlev{\'e} (P) test provided the following
condition among the system parameters (masses, interaction strengths)
is satisfied
\begin{eqnarray} \label{eq:omega12_explicitly}
   a_{12} &=& \frac{(m_{1}a_{11} + m_{2}a_{22})}
                 {M (m_{1} + m_{2})} \pm \hspace*{40mm} \\
          & & \frac{\sqrt{\strut(m_{1}a_{11} + m_{2}a_{22})^{2} +
                                     4M(M-2) m_{1}m_{2} a_{11} a_{22}}}%
                       {M (m_{1} + m_{2})}
   \nonumber
\end{eqnarray}
with $M \equiv \left \lbrack \left ( 2m + 1 \right )^{2} + 7
\right \rbrack /16$, where $m$ can be treated as an arbitrary
positive integer number ($m = 0,1,2,\ldots$). This expression
depends only on the ratios $m_{1}/m_{2}$, $a_{11}/a_{12}$, and
$a_{12}/ a_{22}$ involving the characteristic parameters of the
BEC, and the external potential families can be classified by the
different values of $m$. Experimentally preferred spatially
harmonic ($\sim x^{2}$) potential-families fall into the category
$m=2$ ($M=2$). The condition given by
equation~(\ref{eq:omega12_explicitly}) can be written in the
following interesting form $ M=A_{22} f_{D2}(a_{12}) -
A_{11}f_{B1}(a_{12})$. This relation connects two independent
analysis, namely the solitonic stability conditions
(\ref{eq:ranges}-b) and the P test encoded into
equation~(\ref{eq:omega12_explicitly}). The left hand side is
responsible for the fulfillment of the P test which constitutes
one of the ingredients of integrability conditions
\cite{Ablowitz1991} meaning among many others that solitonic
solution of the GP equations can be found by the method of inverse
scattering transform. Right hand side contains the functions
$f_{B1}$ and $f_{D2}$ which should be positive in order that the
coupled GP equations may have static solitonic solutions of BD or
DB types.

\section{Results and discussion%
         \label{sec:results}}

\subsection{$a_{12}$ values allowing stability and BD solitonic excitations of BECs%
            \label{subsec:stability_of_our_model}}

In this subsection the BD stability ranges of $a_{12}$ will be
presented in tabulated form, using both triplet and singlet
intraspecies data as input, for ninety pairs of alkali atoms. Note
that the constraint arising from TF stability is not included in the
data given.

Tables~\ref{table:triplet_ranges_for_a12} and
\ref{table:singlet_ranges_for_a12} give, separated from each other by
semicolons, the limiting point values of the intervals of the
interspecies scattering length $a_{12}$ satisfying the stability
conditions encoded into inequalities (\ref{eq:ranges}-b). These
conditions express the possibility of developing solitonic
excitations of static BD or DB type from the ground state of the
two-component BECs, irrespective of the particle numbers $N_{1}$ and
$N_{2}$ of the mixture. The data have been calculated, respectively,
with triplet and singlet intraspecies scattering lengths $a_{ii}$
($i=1,2$) listed in the columns headed by the labels
$a_{\mbox{\footnotesize{t}}}$ and $a_{\mbox{\footnotesize{s}}}$. The
exception is the $^{133}$Cs atom where only the average scattering
length ($a_{ii} = + \sqrt{\sigma_{exp}/8\pi}$) has been available
(therefore the parenthesis).

In general we observe in tables~\ref{table:triplet_ranges_for_a12}
and \ref{table:singlet_ranges_for_a12} that one of the limiting
points of the intervals of the corresponding DB and BD cases
coincides; a fact easily derivable by using the relation $0 =
f_{B1}(a_{12}) = f_{D2}(a_{12})$. The diagonal elements of
tables~\ref{table:triplet_ranges_for_a12} and
\ref{table:singlet_ranges_for_a12} represent one-component BECs which
is out of scope of the present interest so that the data are missing
there. Furthermore, the rows corresponding to $^{7}$Li, $^{39}$K, and
$^{85}$Rb of table~\ref{table:triplet_ranges_for_a12} do not contain
any results as well, if their partner owns a positive triplet
intraspecies scattering length. The lack of information means here
that we do not find any region of triplet interspecies scattering
length $a_{12}$ which gives rise to a stable DB type excitation. On
the other hand all items of the column headed by $^{7}$Li, $^{39}$K,
and $^{85}$Rb are filled by data. This means that there are definite
ranges of triplet $a_{12}$ which support BECs with excitations of BD
type where the $^{7}$Li, $^{39}$K, and $^{85}$Rb atoms play the role
of the B component (owing to their attractive interaction character).
The above argumentation can be confirmed by using stability
conditions (\ref{eq:ranges}-b) with $a_{11}<0$ and $a_{22}>0$.

In the case when both $a_{11}$ and $a_{22}$ are positive we observe
the existence of BECs with both DB or BD formations within a finite
interval of positive interspecies scattering length $a_{12}$. This
can be understood physically in the following way. In the context of
the mean field theory, an effective potential is created by the
$i$-th atoms acting repulsively among each other (positive scattering
lengths) and thus forming the cavity of the D component. Inside this
cavity there is a possibility for the $j$-th atoms to form a bunch
representing the B component despite their inherent repulsive
character.

The values of $a_{12}$ given in tables
\ref{table:triplet_ranges_for_a12} and
\ref{table:singlet_ranges_for_a12} can be used for orientation and
in design of two-component BECs. One starts from known
intraspecies scattering lengths $a_{ii}$ ($i=1,2$) which are
represented by the data denoted by $a_{t}$ or $a_{s}$ in the
tables. Then one performs the range determination by using
inequalities~(\ref{eq:ranges}-b). Now one is provided with guessed
values of interspecies scattering length $a_{12}$ which can
maintain stable two-component BEC configuration. If, from some
other sources, we are assured that the actual physical values of
$a_{12}$ falls outside the range calculated then one may try
either to move the $a_{ij}$ values into the proper direction by
utilizing the Feshbach-resonance procedure~\cite{Cubizolles2003}
or to carry out BB or DD stability analysis to learn what type (if
any) of static solitonic excitations of the components are
possible (suggesting, at the same time, stability of the ground
state configurations as well).

However, if all $a_{ij}$'s ($i,j=1,2$) satisfy
inequalities~(\ref{eq:ranges}-b), then one may start to carry out the
production of two-component BEC using information obtained by fine
tuning the component numbers $N_{i}$ as described in Ref.
\cite{Apagyi2004}.

\subsection{$a_{12}$ from earlier studies}
\label{subsec:earlier_studies}

There are very few interspecies scattering length $a_{12}$ data in
the literature. The cases studied so far include the pairs composed
of different isotopes of rubidium atoms \cite{Burke1999}, of those of
potassium atoms \cite{SL39_41K}, of potassium and rubidium atoms
\cite{Ferrari2002, Simoni2003}, of cesium and rubidium
atoms~\cite{Jamieson2003}, and of lithium and cesium atoms
\cite{Mosk2001}. In the latter case the sign of $a_{12}$ remains
undetermined since we can derive, by using the relation
$\sigma_{\mbox{\footnotesize{el}}} = 4 \pi a_{12}^{2}$, only the
absolute value of the interspecies scattering length from the
measured unpolarized elastic cross-section data
$\sigma_{\mbox{\footnotesize{el}}} (^7\mathrm{Li} -
^{133}\mathrm{Cs}) \approx 5\cdot 10^{-12}$ \,cm$^{2}$ to be
$a_{12}(^7\mathrm{Li}-^{133}\mathrm{Cs})=\pm 6.3$ nm. This value of
$a_{12}$ is to be compared to our ranges of $a_{12}$ characterized by
the limiting points $-0.1;4.6$ and $0.9;4.6$\, nm obtained,
respectively, with triplet and singlet intraspecies parameters
$a_{11}=-1.4$ and $1.7$ nm for the lithium atoms, and $a_{22}=2.4$ nm
for the cesium atom (where the latter value also represents
unpolarized data \cite{SL133Cs_1} with known positive sign). From the
comparison we predict the plus sign for the unpolarized interspecies
scattering length, i.e.,
$a_{12}(^7\mathrm{Li}-^{133}\mathrm{Cs})=+6.3$ nm, provided the
stability can be attained.

For the potassium-rubidium system we compare our ranges
$a_{12}(^{39}\mathrm{K}-^{87}\mathrm{Rb})=-0.6;7.6$ nm and
$a_{12}(^{41}\mathrm{K}-^{87}\mathrm{Rb})=4.1;7.5$ nm obtained using
triplet intraspecies data (i.e. with $a_{11}(^{39}\mathrm{K})=-0.9$
nm, $a_{11}(^{41}\mathrm{K})=3.4$ nm, and
$a_{22}(^{87}\mathrm{Rb})=5.5$ nm) to the triplet interspecies values
$a_{12}(^{39}\mathrm{K}-^{87}\mathrm{Rb})=1.6$ nm \cite{Ferrari2002}
and $a_{12}(^{41}\mathrm{K}-^{87}\mathrm{Rb})=8.1^{+0.5}_{0.3}$ nm
given in Ref. \cite{Simoni2003}. We see here that our findings are
comparable with the calculated values of $a_{12}$.

For the cesium-rubidium system a recent work~\cite{Jamieson2003}
lists various $a_{12}$ values calculated using six different
ab-initio potential for both triplet and singlet collisions. The
calculated data lying between $a_{12}=2;19$ nm for the singlet and
between $a_{12}=-8.5;3.2$ nm for the triplet $^{133}$Cs-$^{85}$Rb
system compare well to ours given by the BD ranges of $a_{12}
(^{133}\mathrm{Cs}-^{85}\mathrm{Rb})=16.9$; $97$ nm and
$a_{12}(^{85}\mathrm{Rb}-^{133}\mathrm{Cs})=3.0$; $16.9$\, nm for the
singlet and $a_{12}(^{85}\mathrm{Rb}-^{133}\mathrm{Cs})=-14.8$; $2.9$
nm for the triplet arrangement, respectively. In the case of the
$^{87}$Rb-$^{133}$Cs DB system the singlet scattering lengths are
comparable; for example, the calculated singlet value of $a_{12}=3.0$
nm of the ab-initio result VI~\cite{Jamieson2003} compares well to
our range of $a_{12}(^{87}\mathrm{Rb}-^{133}\mathrm{Cs})=2.9$; $3.3$
nm.

Finally, we mention the singlet interspecies scattering length for
different isotopes of the potassium atoms, namely the value
$a_{12}(^{39}\mathrm{K}-^{41}\mathrm{Kb})=5.9$ nm \cite{SL39_41K} to
be compared to our range
$a_{12}(^{39}\mathrm{K}-^{41}\mathrm{K})=5.7$; $7.0$ nm of the DB
case. Again, a clear compatibility is observed.

We may thus conclude that the solitonic excitation method carried out
in section \ref{sec:TheoreticalBackground} is capable to assess
approximate intervals of interspecies scattering length $a_{12}$
which may encompass the actual values themselves.

\subsection{$a_{12}$ values derived from Painlev{\'e} test}
   \label{subsec:comparison_to_Painleve}

In table~\ref{table:m_values} the values of $a_{12}$ calculated from
equation~(\ref{eq:omega12_explicitly}) are shown for the harmonic
oscillator potential obtained with singlet (upper diagonal) and
triplet (lower diagonal) intraspecies scattering lengths $a_{ii}$,
$i=1,2$. Although $a_{12}=0$ is always a solution of
equation~(\ref{eq:omega12_explicitly}) at $M=2$, this value is not
shown in table~\ref{table:m_values} because it represents the
uncoupled case.

Since the P test of a nonlinear partial differential equation
primarily serves to explore its singularity structure, it is only
implicitly connected with soliton formation. Therefore we do not in
general expect that all values of table~\ref{table:m_values} match
with those encompassed by the $a_{12}$ intervals of
table~\ref{table:triplet_ranges_for_a12} and
\ref{table:singlet_ranges_for_a12}. Yet, the values of items in
tables~\ref{table:triplet_ranges_for_a12} and \ref{table:m_values}
(lower diagonal) and tables~\ref{table:singlet_ranges_for_a12} and
\ref{table:m_values} (upper diagonal) compare well. From this
comparison we can learn that only those DB excitations are possible
in which the D component is represented by the heavier elements
listed in the first line of table~\ref{table:m_values}. Exceptions
are the $^{7}$Li atom which always plays the role of the B component,
and the $^{85}$Rb atom which always constitutes the role of the D
component (when singlet scattering is considered). This result is
physically reasonable, since in a mixture of atoms with different
masses and comparable (positive) interspecies scattering lengths, the
atoms of the slower (more massive) component is likely to form a dip
of D soliton, into which the atoms of the swifter (lighter) component
can be captured to create a B soliton. The exceptions occur if the
repulsion/attraction between like atoms is strong enough as in the
case of $^{85}$Rb/$^{7}$Li to cause a situation with permanent D/B
components.

\section{Summary}
   \label{sec:summary_of_the_method}

The assessment of the interspecies interaction parameter $a_{12}$ at
which two-component atomic BEC exhibits stable configuration is an
important task, and can be carried out with the present method in the
following way.

In principle, all the ground state parameters including $a_{12}$ can
be extracted from the TF equilibrium density profiles. This method is
rather ambiguous and, therefore, restricting the number of uncertain
parameters or their ranges may prove useful in design of
two-component BECs. Moreover, in this fitting procedure the particle
numbers are not coupled, they can be chosen independently of each
other. The coupling is represented mostly by the inter- and
intraspecies scattering length parameters $a_{ij}$ ($i,j=1,2$) being
strongly correlated with each other.

If one is interested in producing excitations of BEC in a form of
static solitons then one first determines a region of interspecies
scattering length $a_{12}$ which may support the soliton formation.
This task can be performed by simultaneous solution of the
inequalities (\ref{eq:ranges}-b) for $a_{12}$, provided the
interaction parameters $a_{11}$ and $a_{22}$ are given. This step is
again free of the particle number parameters $N_{1}$ and $N_{2}$ and
gives the broadest region of $a_{12}$ for which DB or BD static
solitons ever can be created inside the stable two-component BEC. For
the two examples treated explicitly, table
\ref{table:triplet_ranges_for_a12} contains such intervals between
$a_{12}=-0.6:6.1$ nm for the $^{7}$Li-$^{23}$Na triplet system, and
$a_{12}=4.1;7.5$ nm in the case of the $^{41}$K-$^{87}$Rb triplet
scattering.

If the actual value of intraspecies scattering length $a_{12}$
{\emph{which must be known from other source}}, is within the
interval determined above, then one may perform a fine tuning for the
particle number ratio $N_{2}/N_1$ at which the solitonic excitation
may be implemented. This fine tuning is accomplished by solving the
coupled nonlinear equations (11) of Reference \cite{Apagyi2004} for
the ratio $N_{2}/N_{1}$ and the size parameter $k$ of the solitons.
For the two explicit examples chosen, figures~\ref{fig:bd_amplitudes}
contain the results with $N_{2}$ kept fixed.

Bright-dark soliton combinations in one- and two-comp\-onent
condensates have already been considered in a variety of papers,
including \cite{Dum1998a, Busch2001a, Ohberg2001a,
Kevrekidis2003}. Experimental creation of both bright
\cite{Khaykovich2002a} and dark \cite{Denschlag2000} solitons have
been observed in case of one-component condensates. We hope that
fine tuning the system parameters (scattering lengths, particle
numbers, trapping frequencies) as outlined above may also help in
realizing coupled solitonic excitations in two-component
Bose-Einstein condensates.

\begin{acknowledgement}
This work has been supported by OTKA under Grant Nos. T038191,
T047035 and M36482.
\end{acknowledgement}

\bibliographystyle{unsrt}
\bibliography{sabiblio}

\newpage


\def\cell#1#2{\vspace*{0mm}
              \parbox[t]{15mm}{
              \scriptsize{#1};\scriptsize{#2}
                      }}
\def\emptycell{\vspace*{0mm}
                      \parbox[t]{12mm}{
                      }}
\def\rulecell{\vspace*{0mm}
                      \parbox[t]{12mm}{
                      \textemdash
                      }}

\newcommand{\rot}[2]{\hspace{2pt}
                     \begin{rotate}{90}
                        \small \hspace{#1} #2
                     \end{rotate}
                     \hspace{-6pt}
                    }

\setlength{\extrarowheight}{2pt} \setlength{\doublerulesep}{0pt}
\begin{sidewaystable}[htb!]
   \centering
   \caption{Intervals of triplet interspecies scattering lengths $a_{12}$
            assessed by static solitonic excitations of two component BECs
            composed of alkalis indicated. The column
            $a_{\mbox{\footnotesize{t}}}$ gives the triplet intraspecies
            scattering length $a_{ii}$ taken from the references:
            $^{1}$H \cite{SL1H} ,
            $^{7}$Li \cite{SL7Li_3, SL7Li_1,SL7Li_2, SL7Li_5, SL7Li_4},
            $^{23}$Na \cite{SL23Na_1, SL23Na_2, SL23Na_3},
            $^{39}$K, $^{41}$K \cite{SL39_41K, SL40_41K},
            $^{83}$Rb, $^{85}$Rb, $^{87}$Rb \cite{SL87Rb_2, Burke1999},
            $^{133}$Cs, $^{135}$Cs \cite{SL133Cs_1, SL133Cs_2, SL133Cs_3}.
            All values are in unit of nm.
            \label{table:triplet_ranges_for_a12}}
    {\footnotesize{
    \begin{tabular}[t]{c>{\hfill \scriptsize $}p{8mm}<{$}>{\scriptsize $}p{8mm}<{$}*{10}{>{$}p{14mm}<{$}}}
    \hline\hline
    & &
    \multicolumn{10}{c}{\textbf{Bright-component}} \\
    \cline{4-13} &
    \hspace*{-3mm} \mbox{\scriptsize{Element}} &
    \hspace*{4mm}  \mbox{\small $a_{\mbox{\footnotesize{t}}}$}
                       & ^{1}{\mathrm{H}}
                       & ^{7}{\mathrm{Li}}
                       & ^{23}{\mathrm{Na}}
                       & ^{39}{\mathrm{K}}
                       & ^{41}{\mathrm{K}}
                       & ^{83}{\mathrm{Rb}}
                       & ^{85}{\mathrm{Rb}}
                       & ^{87}{\mathrm{Rb}}
                       & ^{133}{\mathrm{Cs}}
                       & ^{135}{\mathrm{Cs}}  \\
     \hline\hline
 & ^{1}{\mathbf{H}}    & \hfill 0.1
                       & \emptycell
                       & \cell{-2.4}{0.0}
                       & \cell{0.0}{0.2}
                       & \cell{-1.8}{0.0}
                       & \cell{0.0}{0.2}
                       & \cell{0.0}{0.1}
                       & \cell{-37.6}{0.0}
                       & \cell{0.0}{0.2}
                       & \cell{0.0}{0.1}
                       & \cell{0.0}{0.1} \\
  & ^{7}{\mathbf{Li}}  & \hfill -1.4
                       & \cell{--}{--}
                       & \emptycell
                       & \cell{--}{--}
                       & \cell{-1.5}{-0.8}
                       & \cell{--}{--}
                       & \cell{--}{--}
                       & \cell{-35.1}{-2.7}
                       & \cell{--}{--}
                       & \cell{--}{--}
                       & \cell{--}{--} \\
  & ^{23}{\mathbf{Na}} & \hfill 4.0
                       & \cell{0.2}{7.7}
                       & \cell{-0.6}{6.1}
                       & \emptycell
                       & \cell{-1.1}{3.0}
                       & \cell{2.8}{3.5}
                       & \cell{1.7}{3.4}
                       & \cell{-30}{1.7}
                       & \cell{1.7}{3.8}
                       & \cell{1.2}{2.2}
                       & \cell{1.2}{3.8} \\
  & ^{39}{\mathbf{K}}  & \hfill -0.9
                       & \cell{-0.4}{-0.1}
                       & \cell{--}{--}
                       & \cell{--}{--}
                       & \emptycell
                       & \cell{--}{--}
                       & \cell{--}{--}
                       & \cell{-26}{-3.8}
                       & \cell{--}{--}
                       & \cell{--}{--}
                       & \cell{--}{--} \\
  & ^{41}{\mathbf{K}}  & \hfill 3.4
                       & \cell{0.1}{6.6}
                       & \cell{-0.4}{5.8}
                       & \cell{3.5}{4.4}
                       & \cell{-0.9}{3.5}
                       & \emptycell
                       & \cell{2.3}{3.6}
                       & \cell{-25.4}{2.2}
                       & \cell{2.2}{4.0}
                       & \cell{1.6}{2.4}
                       & \cell{1.6}{4.2} \\
  & ^{83}{\mathbf{Rb}} & \hfill 4.2
                       & \cell{0.1}{8.3}
                       & \cell{-0.2}{7.7}
                       & \cell{3.4}{6.6}
                       & \cell{-0.6}{5.7}
                       & \cell{3.6}{5.6}
                       & \emptycell
                       & \cell{-19.2}{4.1}
                       & \cell{4.1}{4.8}
                       & \cell{3.1}{3.2}
                       & \cell{3.2}{5.3} \\
  & ^{85}{\mathbf{Rb}} & \hfill -19.0
                       & \cell{--}{--}
                       & \cell{-2.7}{-0.2}
                       & \cell{--}{--}
                       & \cell{-3.8}{-0.6}
                       & \cell{--}{--}
                       & \cell{--}{--}
                       & \emptycell
                       & \cell{--}{--}
                       & \cell{--}{--}
                       & \cell{--}{--} \\
  & ^{87}{\mathbf{Rb}} & \hfill 5.5
                       & \cell{0.2}{10.9}
                       & \cell{-0.2}{10.2}
                       & \cell{3.8}{8.7}
                       & \cell{-0.6}{7.6}
                       & \cell{4.1}{7.5}
                       & \cell{4.8}{5.6}
                       & \cell{-18.8}{5.6}
                       & \emptycell
                       & \cell{3.6}{4.4}
                       & \cell{4.3}{6.1} \\
  & ^{133}{\mathbf{Cs}}& \hfill (2.4)
                       & \cell{0.0}{4.8}
                       & \cell{-0.1}{4.6}
                       & \cell{2.2}{4.1}
                       & \cell{-0.4}{3.7}
                       & \cell{2.4}{3.7}
                       & \cell{3.0}{3.1}
                       & \cell{-14.8}{2.9}
                       & \cell{2.9}{3.6}
                       & \emptycell
                       & \cell{2.4}{4.2} \\
  \rot{5mm}{\textbf{Dark-component}}
  & ^{135}{\mathbf{Cs}}
                       & \hfill  7.2
                       & \cell{0.1}{14.3}
                       & \cell{-0.1}{13.7}
                       & \cell{3.8}{12.3}
                       & \cell{-0.4}{11.2}
                       & \cell{4.2}{11.1}
                       & \cell{5.3}{8.9}
                       & \cell{-14.7}{8.9}
                       & \cell{6.1}{8.8}
                       & \cell{4.2}{7.3}
                       & \emptycell \\
\hline\hline
\end{tabular}}}
\end{sidewaystable}

\newpage


\setlength{\extrarowheight}{2pt} \setlength{\doublerulesep}{0pt}
\begin{sidewaystable}[htb!]
   \centering
   \caption{Intervals of singlet interspecies scattering lengths $a_{12}$
            assessed by static solitonic excitations of two component
            BECs composed of alkali pairs indicated. The column
            $a_{\mbox{\footnotesize{s}}}$ gives the singlet scattering
            length $a_{ii}$ of atoms taken from the references:
            $^{7}$Li \cite{SL7Li_3, SL7Li_5, SL6Li, SL7Li_4},
            $^{23}$Na \cite{SL23Na_1, SL23Na_2},
            $^{39}$K, $^{41}$K \cite{SL39_41K},
            $^{83}$Rb, $^{85}$Rb, $^{87}$Rb \cite{Burke1999},
            $^{133}$Cs, $^{135}$Cs \cite{SL133Cs_1, SL133Cs_2}.
            (For $^{1}$H the triplet value has been used \cite{SL1H}).
            All values are in unit of nm.
            \label{table:singlet_ranges_for_a12}}
    {\footnotesize{
    \begin{tabular}[t]{c>{\hfill \scriptsize $}p{8mm}<{$}>{\scriptsize $}p{8mm}<{$}*{10}{>{$}p{14mm}<{$}}}
    \hline\hline
    & &
    \multicolumn{10}{c}{\textbf{Bright-component}} \\
    \cline{4-13} &
    \hspace*{-3mm} \mbox{\scriptsize{Element}} &
    \hspace*{4mm}  \mbox{\small $a_{\mbox{\footnotesize{s}}}$}
                       & ^{1}{\mathrm{H}}
                       & ^{7}{\mathrm{Li}}
                       & ^{23}{\mathrm{Na}}
                       & ^{39}{\mathrm{K}}
                       & ^{41}{\mathrm{K}}
                       & ^{83}{\mathrm{Rb}}
                       & ^{85}{\mathrm{Rb}}
                       & ^{87}{\mathrm{Rb}}
                       & ^{133}{\mathrm{Cs}}
                       & ^{135}{\mathrm{Cs}}  \\
  \hline\hline
 & ^{1}{\mathbf{H}}    & \hfill  (0.1)
                       & \emptycell
                       & \cell{-0.0}{0.2}
                       & \cell{0.0}{0.2}
                       & \cell{0.0}{0.2}
                       & \cell{0.0}{0.2}
                       & \cell{0.0}{0.1}
                       & \cell{0.0}{0.6}
                       & \cell{0.0}{0.1}
                       & \cell{0.0}{0.1}
                       & \cell{0.0}{0.2} \\
  & ^{7}{\mathbf{Li}}  & \hfill  1.7
                       & \cell{0.2}{3.0}
                       & \emptycell
                       & \cell{0.8}{1.8}
                       & \cell{0.5}{2.5}
                       & \cell{0.5}{1.6}
                       & \cell{0.3}{1.3}
                       & \cell{0.3}{7.7}
                       & \cell{0.3}{1.5}
                       & \cell{0.2}{0.9}
                       & \cell{0.2}{2.9} \\
  & ^{23}{\mathbf{Na}} & \hfill 2.7
                       & \cell{0.2}{5.3}
                       & \cell{1.8}{4.2}
                       & \emptycell
                       & \cell{2.0}{4.3}
                       & \cell{1.9}{3.3}
                       & \cell{1.2}{2.5}
                       & \cell{1.2}{15.1}
                       & \cell{1.1}{2.9}
                       & \cell{0.8}{1.8}
                       & \cell{0.8}{6.0} \\
  & ^{39}{\mathbf{K}}  & \hfill  7.3
                       & \cell{0.2}{14.2}
                       & \cell{2.5}{12.4}
                       & \cell{4.3}{9.2}
                       & \emptycell
                       & \cell{5.7}{7.0}
                       & \cell{4.6}{4.7}
                       & \cell{4.6}{28.0}
                       & \cell{4.5}{5.4}
                       & \cell{3.3}{3.5}
                       & \cell{3.3}{11.5} \\
  & ^{41}{\mathbf{K}}  & \hfill 4.4
                       & \cell{0.2}{8.6}
                       & \cell{1.2}{5.4}
                       & \cell{3.3}{5.7}
                       & \cell{4.6}{5.7}
                       & \emptycell
                       & \cell{3.0}{3.7}
                       & \cell{2.9}{22.1}
                       & \cell{2.9}{4.3}
                       & \cell{2.1}{2.8}
                       & \cell{2.0}{9.0} \\
  & ^{83}{\mathbf{Rb}} & \hfill 3.4
                       & \cell{0.1}{6.8}
                       & \cell{1.3}{6.3}
                       & \cell{2.5}{5.4}
                       & \cell{4.6}{4.7}
                       & \cell{3.7}{4.6}
                       & \emptycell
                       & \cell{3.4}{20.7}
                       & \cell{3.3}{4.0}
                       & \cell{2.6}{2.8}
                       & \cell{2.6}{9.2} \\
  & ^{85}{\mathbf{Rb}} & \hfill 124.8
                       & \cell{0.6}{246}
                       & \cell{7.7}{230}
                       & \cell{15.1}{196}
                       & \cell{28.0}{171}
                       & \cell{22.1}{167}
                       & \cell{20.7}{126}
                       & \emptycell
                       & \cell{24.2}{123}
                       & \cell{16.9}{97}
                       & \cell{55.5}{96} \\
  & ^{87}{\mathbf{Rb}} & \hfill 4.7
                       & \cell{0.1}{9.2}
                       & \cell{1.5}{8.7}
                       & \cell{2.9}{7.4}
                       & \cell{5.4}{6.5}
                       & \cell{4.3}{6.3}
                       & \cell{4.0}{4.8}
                       & \cell{4.7}{24.2}
                       & \emptycell
                       & \cell{3.3}{3.8}
                       & \cell{3.7}{10.8} \\
  & ^{133}{\mathbf{Cs}}& \hfill (2.4)
                       & \cell{0.1}{4.8}
                       & \cell{0.9}{4.6}
                       & \cell{1.8}{4.0}
                       & \cell{3.5}{3.7}
                       & \cell{2.8}{3.7}
                       & \cell{2.8}{3.0}
                       & \cell{3.0}{16.9}
                       & \cell{2.9}{3.3}
                       & \emptycell
                       & \cell{2.7}{7.9} \\
  \rot{7mm}{\textbf{Dark-component}}
  & ^{135}{\mathbf{Cs}}
                       & \hfill 26.0
                       & \cell{0.2}{51.6}
                       & \cell{2.9}{49.5}
                       & \cell{6.0}{44.4}
                       & \cell{11.5}{40.3}
                       & \cell{12.1}{39.9}
                       & \cell{9.1}{32.3}
                       & \cell{32.0}{55.5}
                       & \cell{10.8}{31.6}
                       & \cell{7.9}{26.2}
                       & \emptycell \\
\hline\hline
\end{tabular}}}
\end{sidewaystable}

\newpage

\def\mcell#1{\vspace*{0mm}
                      \parbox[t]{8mm}{\hfill
                      \scriptsize{#1}
                      \,
                      }
              }
\def\m1cell#1{\vspace*{0mm}
                      \parbox[t]{10mm}{\hfill
                      \scriptsize{#1}
                      \,
                      }
              }
\setlength{\extrarowheight}{2pt} \setlength{\doublerulesep}{0pt}
\begin{sidewaystable}[htb!]
   \centering
   \caption{Values of interspecies scattering length $a_{12}$ solving
            the P test condition, equation~(\ref{eq:omega12_explicitly}),
            at classification numbers $m=2$ preferring harmonic oscillator
            trapping potentials. Upper diagonal contains results obtained
            with singlet intraspecies scattering lengths $a_{\mbox{\footnotesize{s}}}$
            listed in the second column, lower diagonal exhibits those with triplet
            $a_{ii}$'s given in the third column as $a_{\mbox{\footnotesize{t}}}$.
            All values are in unit of nm.
            \label{table:m_values}}
   {\footnotesize{
   \begin{tabular}[t]{>{\scriptsize $}p{8mm}<{$}>{\scriptsize $}p{10mm}<{$}>{\scriptsize $}p{10mm}<{$}*{10}{>{$}p{12mm}<{$}}}
   \hline\hline
   \hspace*{-1mm} \mbox{\scriptsize{Element}} &
   \hfill\mbox{\small{$a_{\mbox{\footnotesize{s}}}$}} &
   \hfill\mbox{\small{$a_{\mbox{\footnotesize{t}}}$}}
                         & \, ^{1}{\mathbf{H}}
                         & \, ^{7}{\mathbf{Li}}
                         & \, ^{23}{\mathbf{Na}}
                         & \, ^{39}{\mathbf{K}}
                         & \, ^{41}{\mathbf{K}}
                         & \, ^{83}{\mathbf{Rb}}
                         & \, ^{85}{\mathbf{Rb}}
                         & \, ^{87}{\mathbf{Rb}}
                         &   ^{133}{\mathbf{Cs}}
                         &   ^{135}{\mathbf{Cs}}   \\
   \hline\hline
   ^{1}{\mathbf{H}}     & \hfill (0.1)
                        & \hfill 0.1
                        & \emptycell
                        & \mcell{-1.3}
                        & \mcell{2.6}
                        & \mcell{7.1}
                        & \mcell{4.3}
                        & \mcell{3.4}
                        & \mcell{123.3}
                        & \mcell{4.6}
                        & \mcell{2.4}
                        & \mcell{25.8} \\
   ^{7}{\mathbf{Li}}    & \hfill  1.7
                        & \hfill -1.4
                        & \m1cell{-1.2}
                        & \emptycell
                        & \mcell{2.5}
                        & \mcell{6.4}
                        & \mcell{2.9}
                        & \mcell{3.3}
                        & \mcell{115.6}
                        & \mcell{4.5}
                        & \mcell{2.3}
                        & \mcell{24.8} \\
   ^{23}{\mathbf{Na}}   & \hfill 2.75
                        & \hfill 4.0
                        & \m1cell{3.8}
                        & \m1cell{2.9}
                        & \emptycell
                        & \mcell{5.6}
                        & \mcell{3.8}
                        & \mcell{3.3}
                        & \mcell{196.7}
                        & \mcell{4.3}
                        & \mcell{2.5}
                        & \mcell{22.6} \\
   ^{39}{\mathbf{K}}    & \hfill  7.28
                        & \hfill -0.9
                        & \m1cell{-0.9}
                        & \m1cell{-0.9}
                        & \m1cell{0.9}
                        & \emptycell
                        & \mcell{5.8}
                        & \mcell{4.7}
                        & \mcell{87.9}
                        & \mcell{5.5}
                        & \mcell{3.5}
                        & \mcell{21.8} \\
   ^{41}{\mathbf{K}}    & \hfill 3.12
                        & \hfill 3.4
                        & \m1cell{3.3}
                        & \m1cell{2.8}
                        & \m1cell{3.6}
                        & \m1cell{1.3}
                        & \emptycell
                        & \mcell{3.8}
                        & \mcell{85.0}
                        & \mcell{4.6}
                        & \mcell{2.6}
                        & \mcell{20.6} \\
   ^{83}{\mathbf{Rb}}   & \hfill 3.43
                        & \hfill 4.2
                        & \m1cell{4.2}
                        & \m1cell{3.8}
                        & \m1cell{4.1}
                        & \m1cell{2.6}
                        & \m1cell{3.9}
                        & \emptycell
                        & \mcell{64.8}
                        & \mcell{4.1}
                        & \mcell{2.8}
                        & \mcell{17.4} \\
   ^{85}{\mathbf{Rb}}   & \hfill 124.8
                        & \hfill -19.0
                        & \m1cell{-18.8}
                        & \m1cell{-17.6}
                        & \m1cell{-14.1}
                        & \m1cell{-13.3}
                        & \m1cell{-11.6}
                        & \m1cell{-7.5}
                        & \emptycell
                        & \mcell{64.0}
                        & \mcell{50.3}
                        & \mcell{64.3} \\
   ^{87}{\mathbf{Rb}}   & \hfill 4.68
                        & \hfill 5.5
                        & \m1cell{5.4}
                        & \m1cell{5.0}
                        & \m1cell{5.2}
                        & \m1cell{3.5}
                        & \m1cell{4.8}
                        & \m1cell{4.8}
                        & \m1cell{-6.6}
                        & \emptycell
                        & \mcell{3.3}
                        & \mcell{17.6} \\
   ^{133}{\mathbf{Cs}}  & \hfill (2.4)
                        & \hfill (2.4)
                        & \m1cell{2.4}
                        & \m1cell{2.2}
                        & \m1cell{2.6}
                        & \m1cell{1.6}
                        & \m1cell{2.6}
                        & \m1cell{3.1}
                        & \m1cell{-6.0}
                        & \m1cell{3.6}
                        & \emptycell
                        & \mcell{4.8} \\
   ^{135}{\mathbf{Cs}}  & \hfill 26.0
                        & \hfill 7.2
                        & \m1cell{0.7}
                        & \m1cell{6.8}
                        & \m1cell{6.7}
                        & \m1cell{5.4}
                        & \m1cell{6.3}
                        & \m1cell{6.1}
                        & \m1cell{-2.9}
                        & \m1cell{6.5}
                        & \m1cell{4.8}
                        & \emptycell \\
   \hline\hline
   \end{tabular}
   \vspace*{5cm}}}
\end{sidewaystable}

\clearpage

\end{document}